\documentclass{iau} 
\usepackage{graphicx}

\def\mnras{\textit{MNRAS}}

\def\nar{\textit{New Astron. Revs}}

\title[Establishing the impact of powerful AGN on their host galaxies] %% give here short title %%
{Establishing the impact of powerful AGN on their host galaxies}

\author[C.M. Harrison et~al.]   %% give here short author list %%
{C.M. Harrison$^{1}$, S.J. Molyneux$^{2}$, J. Scholtz$^{3}$ \and M.E. Jarvis$^{4,5,6}$
}

\affiliation{
$^1$School of Mathematics, Statistics and Physics, Newcastle University, Newcastle Upon Tyne, NE1 7RU, United Kindgom\\ email: {\tt christopher.harrison@newcastle.ac.uk} \\ [\affilskip]
$^2$Astrophysics Research Institute, Liverpool John Moores University, 146 Brownlow Hill, Liverpool L3 5RF, UK\\
$^3$Department of Space, Earth and Environment, Chalmers University of Technology, Onsala Space Observatory, SE-43992 Onsala, Sweden\\
$^{4}$Max-Planck Institut f\"ur Astrophysik, Karl-Schwarzschild-Str. 1, 85748 Garching, Germany \\
$^{5}$European Southern Observatory, Karl-Schwarzschild-Str. 2, 85748 Garching, Germany \\
$^{6}$Ludwig Maximilian Universit\"at, Professor-Huber-Platz 2, 80539 Munich, Germany\\
}

\pubyear{2020}
\volume{359}  %% insert here IAU Symposium No.
\jname{Galaxy evolution and feedback across different environments}
\editors{T. Storchi-Bergmann, R. Overzier, W. Forman \& R. Riffel, eds.}

\begin{document}

\maketitle

\begin{abstract}
Establishing the role of active galactic nuclei (AGN) during the formation of galaxies remains one of the greatest challenges of galaxy formation theory. Towards addressing this, we summarise our recent work investigating: (1) the physical drivers of ionised outflows and (2) observational signatures of the impact by jets/outflows on star formation and molecular gas content in AGN host galaxies. We confirm a connection between radio emission and extreme ionised gas kinematics in AGN hosts. Emission-line selected AGN are significantly more likely to exhibit ionised outflows (as traced by the [O~{\sc iii}] emission line) if the projected linear extent of the radio emission is confined within the spectroscopic aperture. Follow-up high resolution radio observations and integral field spectroscopy of 10 luminous Type 2 AGN reveal moderate power, young (or frustrated) jets interacting with the interstellar medium. We find that these sources live in highly star forming and gas rich galaxies. Additionally, by combining ALMA-derived dust maps with integral field spectroscopy for eight host galaxies of $z\approx2$ X-ray AGN, we show that H$\alpha$ emission is an unreliable tracer of star formation. For the five targets with ionised outflows we find no dramatic in-situ shut down of the star formation. Across both of these studies we find that if these AGN do have a negative impact upon their host galaxies, it must be happening on small (unresolved) spatial scales and/or an observable galaxy-wide impact has yet to occur.

\end{abstract}

\firstsection % if your document starts with a section,
\section{Introduction}
Supermassive black holes (with masses $\gtrsim$10$^{6}$\,M$_{\odot}$) reside at the centre of galaxies and they release extraordinary amounts of energy when they grow through mass accretion events. During these growth periods they are identified as active galactic nuclei (AGN) using observations across the electromagnetic spectrum (see review in Alexander \& Hickox 2012). To reproduce realistic galaxy populations,  successful cosmological models of galaxy evolution require that the energy released from AGN heats and/or removes the gas in the host galaxy and/or surrounding halos in the process called ``AGN feedback'' (e.g., Bower et~al. 2006; Schaye et~al. 2015; Pillepich et~al. 2019).

\begin{figure}
\centering
    \includegraphics[width=0.9\columnwidth]{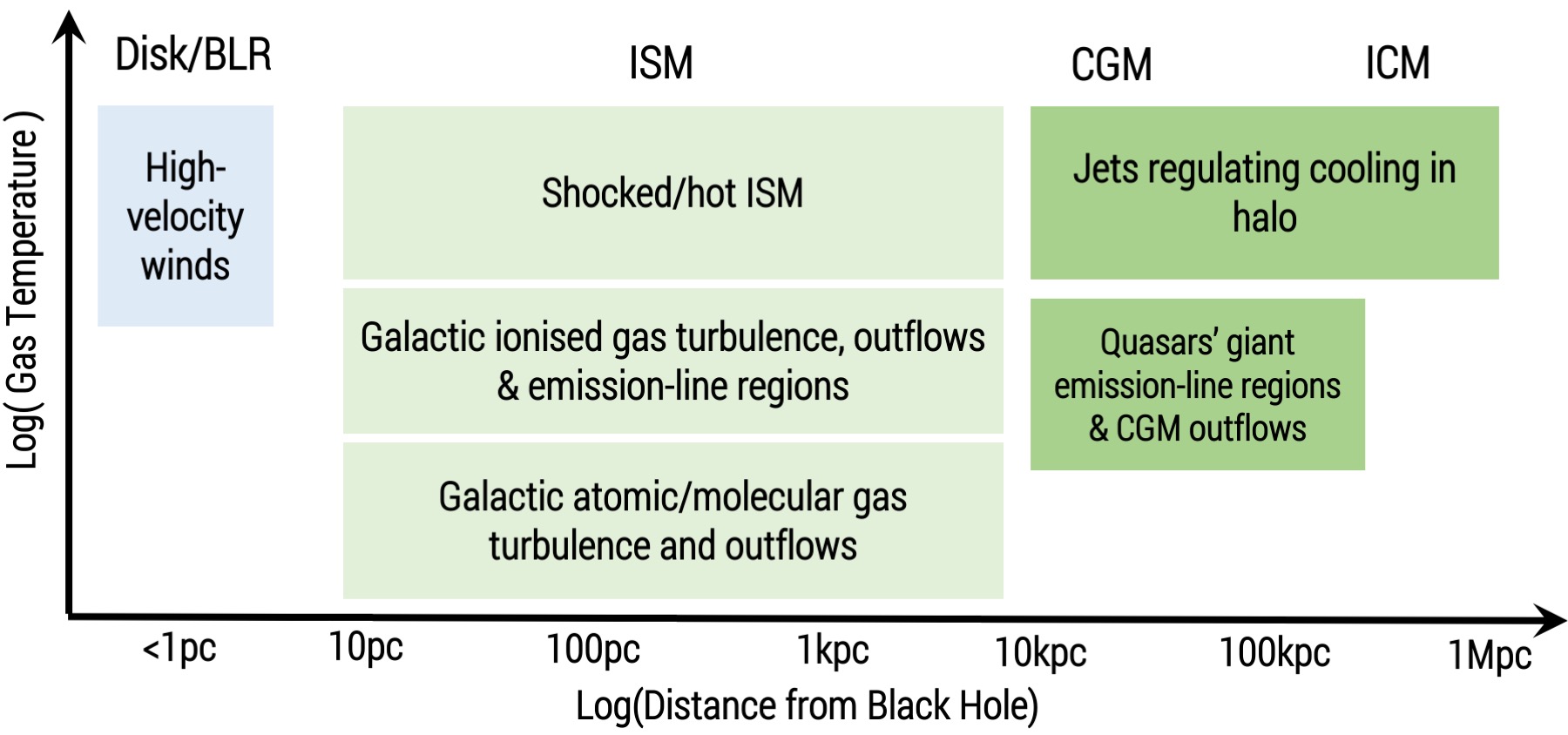}
   \caption{A schematic diagram to summarise the broad observations, that cover a wide range of gas temperatures, that have shown that AGN can inject energy into the gas in and around their host galaxies over a huge range of spatial scales.}
   \label{fig:schematic}
\end{figure}

From an observational perspective, AGN have been seen to interact with gas over a wide range of spatial scales and with gas at a large range of temperatures (see schematic in Fig.~\ref{fig:schematic}). For example: spectroscopic observations of extremely high velocity winds (at a significant fraction of the speed of light) in the regions around the accretion disk or broad line region (BLR; e.g., King \& Pounds 2015); in the interstellar medium (ISM) AGN-driven outflows have been observed across a huge range of temperatures from cold molecular gas to hot X-ray gas (e.g., Cicone et~al. 2018); in the circumgalactic medium (CGM) outflows are also observed and powerful AGN (i.e., quasars) are able to ionise gas out to $\approx$100\,kpc (e.g., Arrigoni Battaia et~al. 2019) and; on the scales of the intracluster medium (ICM) powerful radio jets are seen to be regulating gas cooling, at least in the densest environments (e.g., McNamara \& Nulsen 2012). Despite these efforts there remains little consensus on many key aspects of the physical processes behind, and impact of, AGN feedback. In this article we focus on two key questions: (1) what physical process is most important for AGN to inject energy into the ISM (e.g., radiation, collimated jets of particles, or wide-angle disk winds; see Wylezalek \& Morganti 2018) and; (2) is there observational evidence that AGN directly impact upon the molecular gas or star formation in their host galaxies (e.g., Harrison 2017; Cresci \& Maiolino 2018)? 

\section{The radio--ionised outflow connection}

Previous studies have shown a link between the prevalence and/or velocities of ionised outflows (predominantly traced using the [O~{\sc iii}] emission line) and the radio luminosity of AGN (e.g., Mullaney et al. 2013, Villar-Martin et al. 2014, Zakamska \& Greene 2014). As the basis for our work, Mullaney et al. (2013) used a sample of 24264 optical emission line selected $z<0.4$ AGN from the Sloan Digital Sky Survey (SDSS). They used a combination of stacking analyses and multi-component fits to the emission-line profiles and found that there is a higher prevalence of the most powerful outflows when the radio luminosity is higher. Specifically, AGN of moderate radio luminosity ($L_{\rm 1.4 GHz}$ =$10^{23}$ - $10^{25}$ W Hz$^{-1}$) were found to have the broadest [O~{\sc iii}] profiles and AGN with $L_{\rm 1.4 GHz}$ \textgreater 10$^{23}$ W Hz$^{-1}$ are $\approx$5 times more likely to have extremely broad [O~{\sc iii}] lines (i.e., FWHM$_{\rm Avg}$ \textgreater 1000\,kms$^{-1}$)\footnote{FWHM$_{\rm Avg}$ is the flux-weighted average full-width-half maximum of the two individual Gaussian components that were fitted to the [O~{\sc iii}] emission-line profiles.} compared to lower radio luminous AGN. However, some other similar statistical works have claimed little-to-no correspondence between the radio emission and ionised outflow properties in low redshift AGN (e.g., Woo et~al. 2016; Kauffmann \& Maraston 2019). In the following subsections, we describe how we investigated the radio--outflow connection in greater detail by including spatial information.  

\subsection{Extreme ionised outflows are more prevalent when the radio emission is compact}
In Molyneux et. al (2019) we conducted a study into the relationship between the size of the radio emission (i.e., projected linear spatial extent) and the prevalence of extreme ionised outflows, as traced by [O~{\sc iii}] emission-lines from SDSS spectra. We identified the $\approx$3000 $z<0.2$ AGN from Mullaney et al. (2013), introduced above, with a radio detection in all-sky 1.4\,GHz radio surveys (for more details see Molyneux et. al 2019).

\begin{figure}
\centering
    \includegraphics[width=0.9\columnwidth]{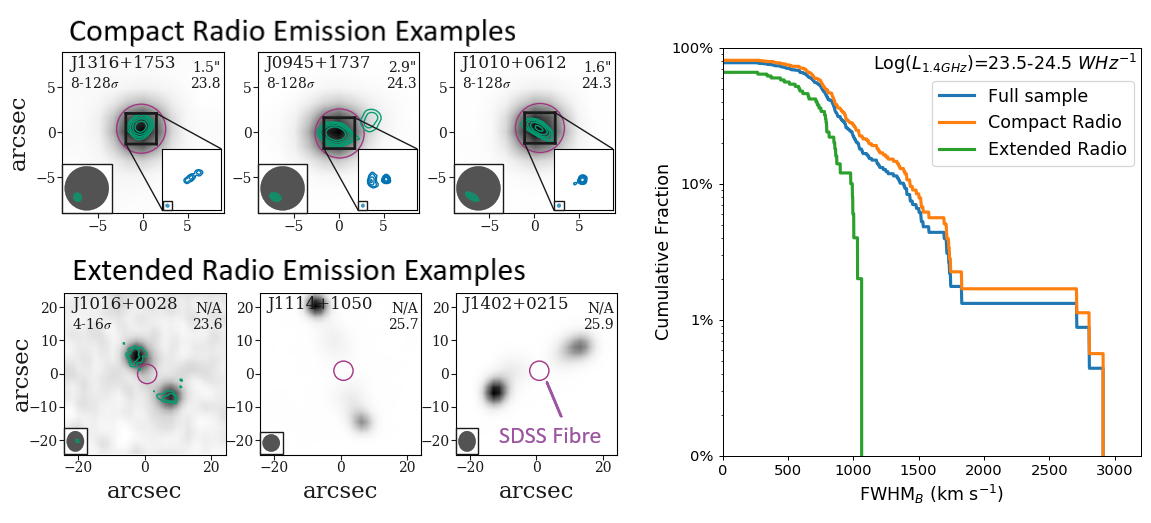}
   \caption{{\em Left:} Example 1.4\,GHz images ($\approx$5\,arcsec resolution) and, where available, overlaid green contours from $\approx$1\,arcsec resolution 1.4\,GHz images. The insets show $\approx$0.3\,arcsec resolution 6\,GHz images (with blue contours). Synthesised beams are represented by appropriately coloured ellipses. Compact radio sources (top row) have their radio emission dominated within the SDSS fibre extent (magenta circles), whilst extended sources (bottom row) show significant emission outside of this. {\em Right:} Cumulative fraction of AGN with [O~{\sc iii}] emission-line components with FWHM$_{B}$ greater than a given value, for: all sources (blue curve), compact radio sources (orange curve) and extended radio sources (green curve). Extreme ionised gas velocities are more prevalent for compact sources. Figure adapted from Molyneux et~al. (2019).}
   \label{fig:Molyneux19}
\end{figure}

To characterise the extent of the radio emission we combined two different approaches. We used: (1) sizes from simple Gaussian models and; (2) an automated morphological classification scheme ({\sc The FIRST Classifier}; Alhassan et al. 2018). It was necessary to combine both of these approaches because whilst the former method has the advantage of providing a quantitative measure of the radio sizes, and corresponding uncertainties, it has the disadvantage of missing structures that are not well characterised by a single elliptical Gaussian model (such as spatially separated radio lobes). Using these two measures we defined {\em compact radio} sources as those with radio emission concentrated within 3\,arcsec or they were classified as {\em extended radio} sources, otherwise (see Molyneux et al. 2019 for full details). This size is equivalent to $\lesssim$5\,kpc, at the median redshift of the sample, and is therefore of the order of the galaxy sizes. Importantly, this cut-off size corresponds to the aperture size of the SDSS fibres, and allows for the most direct comparison to be made between the radio emission and the emission-line properties  observed in the SDSS spectra (see Fig.~\ref{fig:Molyneux19}). Reassuringly, using our follow-up radio data, with $\approx$0.3--1\,arcsec spatial resolution, of a subset of targets, from the Very Large Array, we were able to confirm our size classification method was successful and identified jet-like structures within $\approx$5\,kpc for the compact radio sources (see Fig.~\ref{fig:Molyneux19}).     
 
We produced cumulative distributions of the velocity widths of the [O~{\sc iii}] emission lines for both the compact and extended radio sources\footnote{We found the same results using both the FWHM of secondary broad Gaussian components (FWHM$_{B}$) and the flux-weighted average line widths of two components (FWHM$_{\rm Avg}$)} and a found significantly higher prevalence of extreme [O~{\sc iii}] line widths in the compact radio sources. We demonstrated that this was driven by the differences in radio sizes and was not due to any differences in the distributions of radio or [O~{\sc iii}] luminosities between the two sub-samples. The higher prevalence of ionised outflows in the compact radio sources was most significant in the radio luminosity range log[L$_{\rm 1.4 GHz}$/W Hz$^{-1}$] = 23.5--24.5 (see Fig.~\ref{fig:Molyneux19}). Indeed, this is the radio luminosity range in which AGN are generally accepted to start to dominate the radio emission at low redshift, whereas at lower luminosities the contribution from star formation to the overall radio emission is greater (e.g., Kimball \& Ivezi{\'c} 2008; Condon et al. 2013). The difference between the radio compact and radio extended sources is quantified by the fact that the prevalence of extreme [O~{\sc iii}] emission-line components (with FWHM$_B$\textgreater1000 kms$^{-1}$) is almost four times higher in the compact radio sources in this luminosity range (see Fig.~2).

Our results confirm that there is a connection between radio emission and the prevalence of extreme ionised gas kinematics (outflows) in AGN host galaxies. Importantly, we find that this difference is most extreme in the luminosity range log[L$_{\rm 1.4 GHz}$/W Hz$^{-1}$] = 23.5--24.5 and we suggest that statistical samples which are not complete in this luminosity range may not find such a strong outflow--radio connection (see discussion in Molyneux et~al. 2019). One explanation for this result is that small scale (young or frustrated) radio jets are strongly interacting with the interstellar medium (also see Holt et al. 2008), whilst larger scale radio jets are depositing their energy on larger scales (at least outside of the area covered by the spectroscopic fibre). Indeed, low-to-moderate radio sources with bright optical emission (i.e., radiatively efficient AGN), may represent a key galaxy evolution phase (e.g., Pierce et~al. 2020). To investigate this further requires a combination of radio imaging and spatially-resolved spectroscopy. 

\subsection{Prevalent jet-ISM interactions in radiatively efficient AGN host galaxies}
\begin{figure}
\centering
    \includegraphics[width=0.7\columnwidth]{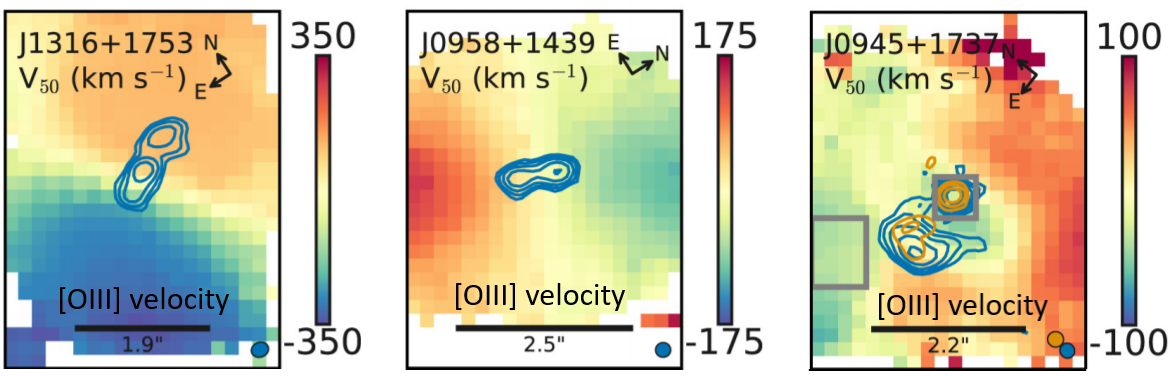}
   \caption{Example radio data and ionised gas velocity maps from integral field spectroscopy for $z\sim0.1$ Type~2 quasars, adapted from Jarvis et~al. (2019). In each case, the radio contours are overlaid (blue and orange) and the corresponding beam(s) are shown in the lower right corner. The scale bar in each image represents 5\,kpc. The ionised gas shows distinct kinematics at the location of the spatially-extended jet/lobe structures indicating jet-ISM interactions.}
   \label{fig:Jarvis19}
\end{figure}

For a more direct investigation into the relationship between ionised outflows and radio emission, in Jarvis et al. (2019), we combined $\approx$0.3--1\,arcsec resolution radio observations with integral field spectroscopy on a sample of 10 Type 2 AGN, selected from the sample in Molyneux et~al. (2019) described above. These sources are in the key radio luminosity range log[L$_{\rm 1.4 GHz}$/W Hz$^{-1}$]=23.5--24.5, are radiatively efficient (with quasar-level bolometric luminosities) and have signatures of outflows in their SDSS spectra (i.e., [O~{\sc iii}] FWHM$_{B}>$700\,km\,s$^{-1}$). Data for three example targets are shown in Figure~\ref{fig:Jarvis19}. In 9/10 of the targets, we confirmed that $\gtrsim$90\% of the radio emission could not be attributed to star formation (see Jarvis et~al. 2019 for details). Furthermore, we found that 80--90\% of these targets show spatially-extended radio structures on 1--25\,kpc scales and these structures are aligned with the gas distribution and/or gas kinematics (e.g., Fig.~\ref{fig:Jarvis19}).

 In Figure~\ref{fig:Jarvis19} we give specific examples of the data. The target J1316+1753 exhibits a strong velocity gradient in [O~{\sc iii}] emission, where the blue and red shifted gas is brightest at the termination of each jet. Similarly J0958+1439 shows a kinematic splitting and co-spatial jets/lobes. The potential jet we observe in J0945+1737, terminates at an [O~{\sc iii}] brightened and blue-shifted gas cloud (indicated by the grey box in the figure). This is evidence of jets hitting a cloud of gas, both pushing the gas away and deflecting the jet. Such observations are signatures of jet-driven outflows (e.g.,  Rosario et al. 2010). 

Overall, our results are consistent with jet–ISM interactions as the cause of the radio--outflow connection seen in statistical samples (e.g., Mullaney et~al. 2013; Molyneux et~al. 2019; Fig.~\ref{fig:Molyneux19}). The combined observations suggest that compact, low-power radio jets, young or frustrated by interactions with the host galaxy ISM, may be responsible for the high-velocity ionised gas, in line with some recent model predictions (e.g., Mukherjee et al. 2018). However, we can not rule out other possible processes, such as nuclear wide-angle winds, that may contribute to producing the radio emission and outflows in the wider sample (e.g., see Zakamska et al. 2016), or in the very nuclear regions of the targets we investigated (see discussion in Jarvis et~al. 2019). A spatially-resolved investigation of the radio--outflow connection, with larger samples, is required to determine how prevalent jet-ISM interactions are across the entire AGN population.

\section{Investigating the impact of AGN on their host galaxies}
The simple {\em identification} of radio jets and AGN-driven ionised outflows does not immediately establish that they have any appreciable {\em impact} on the evolution of their host galaxies (see e.g., Harrison 2017). Towards understanding the possible impact of jets and outflows, we investigated the molecular gas content of the $z\approx0.1$ Type 2 AGN from Jarvis et~al. (2019) (summarised in Section~3.1) and the spatial correlation between outflows and star formation for $z\approx2$ X-ray identified AGN (summarised in Section~3.2). 

\subsection{Outflows and jets associated with high molecular gas fractions at $z\approx0.1$}
\begin{figure}
\centering
    \includegraphics[width=0.5\columnwidth]{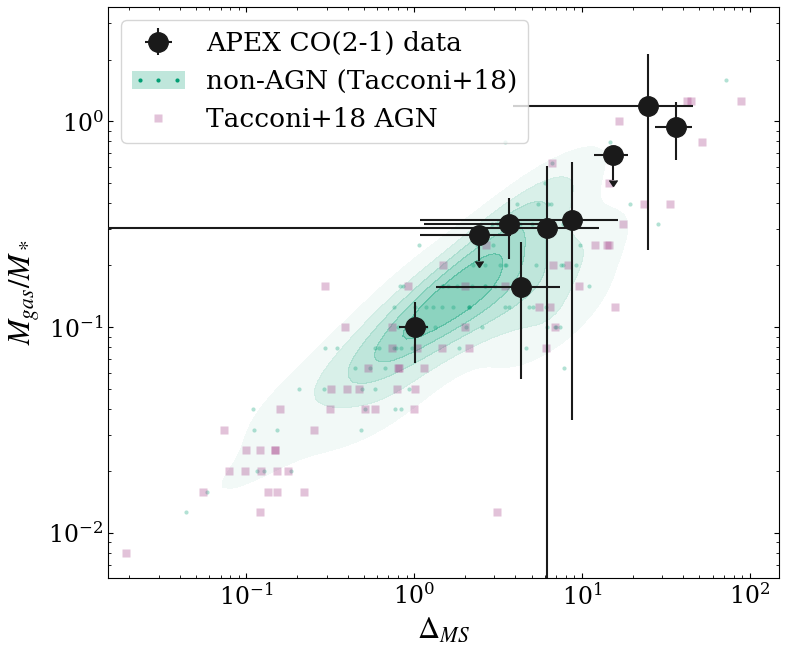}
   \caption{Molecular gas fraction as a function of distance to the main sequence, for our sample (black circles) compared to a redshift-matched comparison sample of galaxies from Tacconi et~al. (2018) (small green circles and contours, with AGN host galaxies marked with pink squares). Our powerful AGN, containing both outflows and jets, follow the overall trends seen in the comparison sample, with moderately higher ($\sim$0.1\,dex) average gas fractions compared to the general star-forming population. Figure adapted from Jarvis et~al. (submitted).}
   \label{fig:Jarvis20}
\end{figure}

For the 9 out of the 10 $z\approx0.1$ AGN in Jarvis et~al. (2019) that showed AGN-dominated radio emission (see Section~2), we used APEX to observe the CO(2--1) emission line (as a tracer of the global molecular gas content; Jarvis et~al., submitted). We found that these targets have CO luminosities consistent with the overall galaxy population for their observed infrared luminosities and once you account for their high specific star formation rates (SFR/$M_{\star}$=0.1--5.9\,Gyr$^{-1}$) and corresponding position relative to the `main sequence' of star forming galaxies ($\Delta_{\rm MS}$; e.g., Sargent et~al. 2014).\footnote{We note that the infrared luminosities we use have been corrected for contamination from the AGN.} Converting these CO luminosities to molecular gas masses implies that the host galaxies have high molecular gas fractions ($M_{gas}/M_{\star}$=0.1--1.2; Fig.~\ref{fig:Jarvis20}; Jarvis et~al., submitted). These gas fractions are consistent with, or moderately higher than, the non-active galaxies with the same specific star formation rates and position relative to the star-forming main sequence (and after carefully applying consistent assumptions) from Tacconi et~al. (2018) (Fig.~\ref{fig:Jarvis20}). Our sources may represent a key phase in galaxy evolution where there are rapid levels of star formation and black hole growth, but the feedback processes we observe (i.e., ionised outflows and associated jets) have not yet been able to have an observable impact upon the global gas content or star formation in their host galaxies (full discussion in Jarvis et~al., submitted).

%%%%%%%%%%%%%%%%%%%%%%%%%%%%%%%%%%%%%%%%%%%%%%%%%%%%%%%%%%%%%%%%%%%%%%%%%%%%%%%%%%%%%%%%%%%%%%%%%%%%%%%%%%%%%%%%%%%%%%%%%%%%%%%%%%%%%%%%%%%%%%%%
% Jan's section 
%%%%%%%%%%%%%%%%%%%%%%%%%%%%%%%%%%%%%%%%%%%%%%%%%%%%%%%%%%%%%%%%%%%%%%%%%%%%%%%%%%%%%%%%%%%%%%%%%%%%%%%%%%%%%%%%%%%%%%%%%%%%%%%%%%%%%%%%%%%%%%%%
\subsection{No dramatic in-situ impact on star formation by ionised outflows at z$\sim$2}

One approach to determine the impact of AGN feedback on star formation, is to use spatially-resolved observations to map both the AGN driven outflows and the star formation in or around the outflows. For example, Cresci et~al. (2015) suggest both {\em suppressed} star formation at the location of an ionised outflow and {\em enhanced} star formation around the edges of the outflow for a $z$=1.6 X-ray identified AGN. Similar findings were presented for three $z$=2.5 extremely powerful (and consequently rare) quasars by Cano-Diaz et~al. (2012) and Carniani et~al. (2016). These works used high-velocity [O~{\sc iii}] emission-line components to map the ionised outflows and narrow (i.e., not associated with the BLR) H$\alpha$ emission to map the spatial distribution of the star formation. 

However, these studies have only mapped the {\em un-obscured} star formation in these AGN host galaxies which is sensitive to potentially missing any dust-obscured star formation. Furthermore, these studies mostly focus on the extremely luminous systems with the most extreme ionised gas outflows. In our pilot study, presented in Scholtz et~al. (2020), we combined integral field spectroscopy (to trace the emissions lines) with ALMA sub-mm interferometry (to trace the dust). In this study we focus on more moderate luminosity, representative, AGN without any pre-selection on the presence of an outflow.

For Scholtz et~al. (2020) we selected z$=$1.5-2.6 AGN from the KASHz (KMOS AGN at High-z) survey (Harrison et~al. 2016), which is an integral field spectroscopic survey of X-ray AGN. The targets were selected to be detected in both H$\alpha$ and [O~{\sc iii}] emission lines and an archival detection in ALMA Band 6 or 7 continuum (corresponding to the rest-frame far-infrared [FIR] emission; $\approx$300-400 $\mu$m; see Scholtz et~al. 2020). Our final sample of eight targets has X-ray luminosities of $10^{43}$-$10^{45.5}$ erg\,s$^{-1}$ and [O~{\sc iii}] emission-line widths of $W_{80}$=300--850 km$s^{-1}$, representative of the parent X-ray AGN population at this redshift (Scholtz et~al. 2020).\footnote{W$_{80}$ is the velocity width containing 80\% of the total emission-line flux.} We established that the rest-frame FIR (ALMA) emission in our targets was tracing star formation heated dust using spectral energy distribution decomposition (i.e., by assessing the possible contribution from AGN emission; Scholtz et~al. 2020). 

\begin{figure}
    \begin{center}\includegraphics[width=9cm]{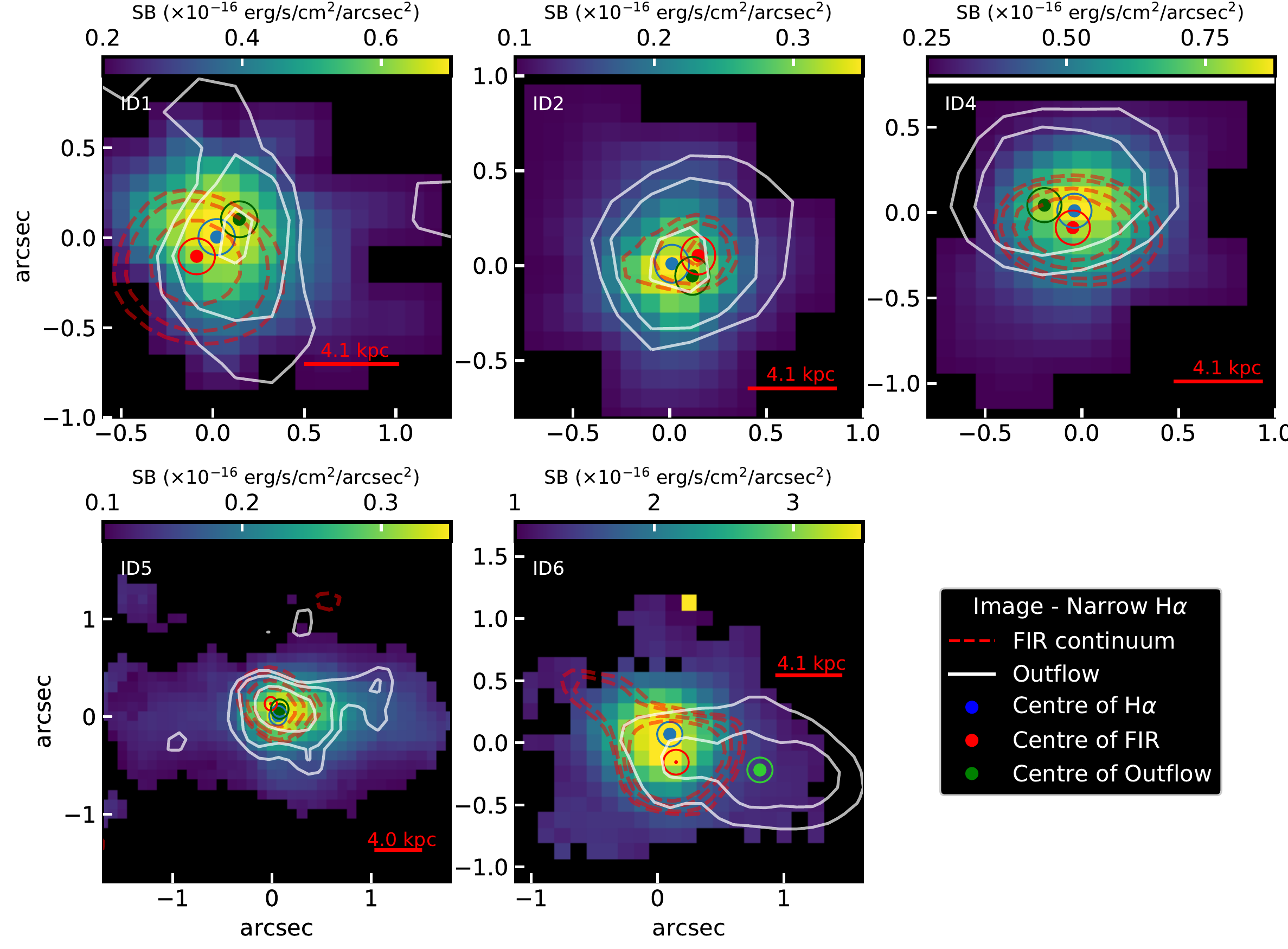}
    \end{center}
   \caption{For the five X-ray AGN, where we identified outflows, we show maps of surface brightness of the narrow line H$\alpha$, with the red-dashed contours showing the distribution of FIR emission (from Scholtz et~al. 2020). White contours show the distribution of ionised outflows (3,4,5 $\sigma$ levels), as defined by the high-velocity wings of the [O~{\sc iii}] emission lines. The blue, red and green points show the peak of the H$\alpha$, FIR and the outflow, respectively. We do not see any significant suppression of the H$\alpha$ emission or the dust emission at the location of the outflows. Figure taken from Scholtz et~al. (2020).}
   \label{fig:SF_outflow}
\end{figure}

We fitted the datacubes containing H$\alpha$ emission with spectral line models for narrow H$\alpha$, H$\alpha$ originating from the BLR and for the [N~{\sc ii}] doublet. This way, we were able to construct a map of the narrow H$\alpha$ emission, without contamination. Furthermore, we detected [O~{\sc iii}] emission-line outflows (i.e., asymmetric emission-line profiles) in five of the eight targets (which is consistent with the $\approx$60\% outflow detection fraction in this luminosity range; Harrison et~al. 2016). For these five targets \ref{fig:SF_outflow} we display the resulting maps of the narrow H$\alpha$ and the spatial distributions of the ionised outflows (defined as the blue wings of the [O~{\sc iii}] emission line; see Scholtz et~al. 2020) in Figure~\ref{fig:SF_outflow}. The red contours in Figure \ref{fig:SF_outflow} represent the spatial distribution of the dust continuum emission (established from the ALMA data). We note that particular care was taken to align the various datasets by first ensuring a consistent astrometric solution (Scholtz et~al. 2020). 

Across the full sample of eight targets we found that the H$\alpha$ emission underestimated the total global star formation rates, even before accounting for the strong contribution from AGN ionisation to the H$\alpha$ emission. Furthermore, in half of the sample we found significant offsets (1.4\,kpc on average) between the dust and H$\alpha$ emission (see ID1 in Figure~\ref{fig:SF_outflow} for a clear example). In Figure~\ref{fig:SF_outflow} we find that three of the targets (ID1, ID5 and ID6) show significant [O~{\sc iii}] outflows elongated beyond the central regions. However, we do not see any strong evidence that the outflows suppress the star formation; i.e., either through cavities in the H$\alpha$ emission at the location of the ionised outflows (cf. Cano-Diaz et~al. 2012, Cresci et~al. 2015, Carniani et~al. 2016) or cavities in the rest-frame FIR emission. Interestingly, using  deeper H$\alpha$ data we do not find any anti-correlation between H$\alpha$ and [O~{\sc iii}] outflows in ID 6, the same object that was presented in Cresci et al (2015) as having a H$\alpha$ cavity (see Scholtz et~al. 2020 for further discussion). Based on our work, we do not find any evidence that outflows from moderate luminosity AGN instantaneously influence the in-situ star formation inside their host galaxies, at least on $\gtrsim$4\,kpc scales. 

\section{Concluding remarks}
Ionised outflows are prevalent in AGN host galaxies (e.g., Mullaney et~al. 2013; Harrison et~al. 2016). At least at low redshift ($z<0.2$) the prevalence of these outflows is associated with the radio emission (Fig.~\ref{fig:Molyneux19}; e.g., Mullaney et~al. 2013; Zakamska \& Greene 2014; Molyneux et~al. 2019). Our detailed observations of a subset of the most powerful sources reveal that jet-ISM interactions are common and demonstrate that low power radio jets are an important feedback mechanism, even in radiatively efficient AGN (e.g., quasars; Fig.~\ref{fig:Jarvis19}; Jarvis et~al. 2019). However, the relative role of jets over other processes, such as wind-angle winds, is still to be determined. We find the AGN-driven outflows and jets do not have a dramatic galaxy-wide impact on the the host galaxy molecular gas content (at least for our $z\approx0.1$ sample; Fig.~\ref{fig:Jarvis20}; Jarvis et~al. submitted) or in-situ star formation (at least for our $z\approx2$ sample; Fig.~\ref{fig:SF_outflow}; Scholtz et~al. 2020). However, some impact could be occurring on spatial scales below those to which we are sensitive. Alternatively, the cumulative impact of AGN outflows or jets may regulate future galaxy properties over longer timescales  even if they have little impact on the {\em in-situ} star formation (e.g., see discussion in Harrison 2017; Scholtz et~al. 2018; 2020).

\end{document}